\documentclass{article}

\usepackage{arxiv}

\usepackage[utf8]{inputenc} % allow utf-8 input
\usepackage[T1]{fontenc}    % use 8-bit T1 fonts
\usepackage{hyperref}       % hyperlinks
\usepackage{url}            % simple URL typesetting
\usepackage{amsfonts}       % blackboard math symbols
\usepackage{nicefrac}       % compact symbols for 1/2, etc.
\usepackage{microtype}      % microtypography
\usepackage{dirtytalk}
\usepackage[numbers]{natbib}

\title{Voices of Workers: Why a Worker-Centered Approach to Crowd Work Is Challenging}

\author{
 Caifan Du \\
  School of Information\\
  University of Texas at Austin\\
  Austin, TX\\
  \texttt{cfdu@utexas.edu} \\
   \And
 Matthew Lease \\
  School of Information\\
  University of Texas at Austin\\
  Austin, Tx \\
  \texttt{ml@utexas.edu} \\
}

\begin{document}
\maketitle
\begin{abstract}
How can we better understand the broad, diverse, shifting, and invisible crowd workforce, so that we might better support it? We present findings from online observations and analysis of publicly available postings from a community forum of crowd workers. In particular, we observed recurring tensions between crowd workers and journalists regarding media depictions of crowd work. We found that crowd diversity makes any one-dimensional representation inadequate in addressing the wide-ranging experiences of crowd work. We argue that the scale, diversity, invisibility, and the crowds' resistance to publicity make a worker-centered approach to crowd work particularly challenging, necessitating better understanding the diversity of workers and their lived experiences.
\end{abstract}

% keywords can be removed
%\keywords{First keyword \and Second keyword \and More}

\section{Introduction}
Crowdsourcing platforms such as Amazon Mechanical Turk (AMT) provide online marketplaces for task-based microwork \cite{barr2006ai, irani2015cultural}. AMT in particular has been widely used for research and business purposes \cite{smith_gig_2016}, including tasks for natural language processing, image annotation, online experiments, and user studies (e.g., \cite{callison-burch_creating_2010, nowak_how_2010, grady_crowdsourcing_2010, paolacci_running_2010, kittur_crowdsourcing_2008}). Existing studies have estimated over 100K active workers on AMT and more than 2K working at any given time (e.g.,\cite{difallah_demographics_2018}). 

Two primary classes of users engage in economic exchanges on the platform: {\em Requesters} who distribute tasks and crowd workers who complete tasks for payment. A decade ago, \cite{silberman2010ethics} called for more worker-centered research in this domain. Since then, researchers have studied AMT worker demographics \cite{ipeirotis2010demographics, ross_who_2010}, motivation \cite{antin_social_2012,brewer_why_2016,gray_ghost_2019,kaplan_striving_2018,kaufmann_more_2011,martin_being_2014}, pay rates \cite{chilton_task_2010,ipeirotis2010demographics,horton_labor_2010,mcinnis_taking_2016,silberman2018responsible,hara_data-driven_2018, hara_worker_2019}, working experience \cite{gupta_turk-life_2014,martin_being_2014,gray_crowd_2016,martin_turking_2016,mcinnis_taking_2016}, and their collective actions \cite{bederson_web_2011,salehi_we_2015}. While researchers have employed a wide range of methods to understand AMT workers and their work, there has been a focus for eliciting workers' own perspective and voices, such as observing worker interactions in their online community forums \cite{martin_being_2014,martin_turking_2016}, collecting worker responses and accounts through surveys and interviews \cite{gupta_turk-life_2014, martin_turking_2016, whiting_fair_2019}, and gathering requirements directly from workers when building tools to support \say{Turking}, usually via "HITs" (i.e., unit of work called "Human Intelligence Tasks" on AMT) \cite{irani_turkopticon_2013,whiting_fair_2019, salehi_we_2015}. 

We seek to build on this line of research by further enhancing the understanding of why it can be so challenging to devise effective worker-centered support mechanisms for crowd work. While there certainly have been many valuable works in this vein, detailed reflections on the friction involved are less common. 

Through online observation in AMT worker community forums, we note tension between workers and their media exposure. This tension prompted us to analyze worker interactions and narratives posted in online discussion forums as publicly available content. Our qualitative analysis sheds further light upon the causes and mechanisms that render a worker-centered approach to crowd work difficult. At its core, it is the challenge of effectively eliciting and representing a plurality of diverse workers' voices and perspectives \cite{cooper_representing_1995, haraway_modest_witnesssecond_millenium_1997} that make it difficult to speak for or design interventions that are appreciated by these workers. %We unpack what constitute this difficulty in this article. Since 

Because our findings are centered around crowd workers on AMT, we use the term \say{Turkers} to refer to these workers, consistent with how these workers often identify themselves, as we have observed in their online community forums.

\section{Related Work}
A variety of research has focused on learning about and improving the Turking experience. Researchers who rely on the principles of ethnomethodology have investigated how Turkers arrange their day-to-day work activities in interaction with other geographically distributed workers, Requesters, and the AMT platform as both an organizational entity and a task marketplace \cite{martin_being_2014, martin_turking_2016}. \cite{martin_being_2014} were among the very first to analyze the publicly available conversations in Turkers' online forums. They characterized Turkers as economic actors operating in an online labor market, focusing on workplace issues such as labor relations, wages, and work ethics. In \cite{martin_turking_2016}'s later study, they compared US-based vs.\ India-based Turkers to better understand differences in working experience across geographical boundaries. Similarly, \cite{gupta_turk-life_2014} provides a detailed description of Indian Turkers' working lives, in contrast with that of US-based workers based on surveys and interviews. Some specific aspects of Turking in India include adapting to time zone differences and viewing Turking work as an international professional experience in individual career profiles.

Critical designers \cite{irani_turkopticon_2013} have sought to support Turkers via system building efforts, eliciting Turker voices for design ideation. \cite{irani_turkopticon_2013} conducted an open-ended survey as HIT on AMT, getting input from Turkers for compiling \say{Workers' Bill of Rights}. They asked about daily work concerns and then developed a Requester rating system, \emph{Turkopticon}, based on Turker responses. 
\cite{irani_stories_2016} helped to introduce Turkers into HCI conference venues for their better visibility. They also reflect on their experience of constructing and maintaining the technological system as well as building relations with Turkers. Related to our work, they note that the lack of unity in worker voices in online forums makes it challenging to form collective voices and actions, reflecting the difficulty of worker-centered design and advocacy.

\cite{salehi_we_2015} seeks to raise Turkers' voices by organizing collective actions within the Turker community. With the help of the researchers, Turkers collectively authored action guidelines for academic Requesters and a letter to Amazon that advocated fairness and openness toward Turkers \cite{noauthor_guidelines_2017}. Both efforts are mobilized campaigns following the ideation by the Dynamo Turker community. Other researchers \cite{Mankar17-hcomp,whiting_fair_2019}
have formulated technological interventions to help workers earn a minimum hourly wage. Both research groups also surveyed workers via HITs on AMT, e.g., regarding their expectations for fair pay. More recent studies of Turking experience and designing for supporting Turking dominantly rely on Turkers' online comments and HIT surveys \cite{hanrahan2021expertise, saito2019predicting, toxtli2020meta, flores2020challenges}. 

Prior worker-centered studies range from observation of Turkers' online conversations to surveying and interviewing Turkers directly and using their responses to design and develop supportive tools. In particular, active critical designers seek to \say{interrupt worker invisibility} \cite{irani_turkopticon_2013}, helping them to form voices, be heard, and be seen. As all these works have explored some channels to get Turkers' voices and experiences, it is not surprising that it is often the invisibility of Turking that motivates these studies and affects their interpretations. Similarly, \cite{gray_ghost_2019} characterize invisible crowd work as \say{ghost work} through five years of study. Our work builds upon such past scholarship in seeking to provide explanations for the ways in which invisibility operates to make a work-centered approach to crowd work particularly challenging.

Regarding the common popular press narrative of exploited Turkers, \cite{moss2020ethical} presents a data-driven rebuttal that resonates with our own analysis.

\section{Methods}
Our work was prompted by initial observation in Turkers' discussion forums. We began to frequent several known online gathering spaces of crowd workers -- keeping up with the recent updates and reading through historical discussion threads -- to gain a broader understanding of Turkers' online activities and discourse. Those sites include Turker Nation\footnote{\url{https://www.reddit.com/r/TurkerNation/}}, TurkerView\footnote{\url{https://forum.turkerview.com/}}, and several Subreddits\footnote{\url{https://www.reddit.com/reddits/}} dedicated to discussions about Turking or crowd work of all sorts. 

We began to notice a set of similar discussion threads in Subreddit {\tt r/mturk}\footnote{\url{https://www.reddit.com/r/mturk/}}, an AMT forum (having more than 80K members as of September 2021) in which Turkers discuss their daily Turking experiences. Topics span issues such as how to make sense of posted HITs, how to understand and manage rejections and get payments transferred to a bank account, etc. The Subreddit thus serves as a gathering place for Turkers to help each other navigate the dynamics of AMT work.

The set of discussion threads we focused on involve conversations between self-identified journalists and Turkers. Threads were initiated by a journalist, or occasionally a researcher, asking whether Turkers in the forum would participate in an interview to discuss their Turking experiences, for a news article or research on crowd work. The first reply in these threads frequently conveyed some degree of resistance to being interviewed. We began to read through and interpret these discussion threads as recurring events observed in the field \cite{spradley2016participant}. We also further searched the Subreddit using \say{journalist} or \say{interview} as keywords, and our review of search results identified more such threads. 

As our observation and qualitative analysis went on, we further noticed that community members routinely noted and discussed news media portrayals of them. Many discussion threads stemmed from sharing of a new news story covering Turkers, followed by discussion. As in the threads in which journalists sought Turkers to interview, these discussions also contained Turker reactions to news media publicizing them and their work. Searching through the historical postings about news coverage of Turkers in the Subreddit, we identified another set of 12 threads containing substantial discussions of how Turkers themselves make sense of popular narratives about their images, conditions, and experiences.

Overall, we identified 26 threads posted across the past six years, with the latest posted in the last year. Most are substantial discussions; the shortest has three comments while the longest has eighty-nine. These discussion threads evidence similar events: Turkers' reaction to the potential or ongoing publicity about them. These 26 threads were initiated by 21 registered Reddit user accounts; in 5 cases, an interview request was posted more than once as separate threads, but with different conversations unfolding. Each discussion for a thread involved an almost exclusively different set of community members. Only five user accounts, as we identified, appeared in more than one discussion thread.

We analyzed these discussion threads based on the principles of thematic analysis, without any epistemological commitment preceding, coding each thread inductively \cite{braun_using_2006, braun2012thematic}. We treated each thread as one conversation and paid close attention to the role and stance of each participant involved. We analyzed each comment in the thread and  interpreted it in the context of the conversation. We then memo'ed each comment with interpreted meaning as initial codes. When one comment became sufficiently long to carry complex meanings, we further decomposed it and memo'ed them with multiple codes accordingly. Next, we sought converging themes across these codes and threads. As we organized coded data under different themes, we reviewed the whole coded data set in relation to identified themes to ensure mutual fit. A few codes were dropped in this phase, carrying some special meaning but not connected to any themes, or not supporting a theme-level finding given their sparse presence in the data.

\section{Voices of Workers}
\label{sec:voices}

In this section, we discuss our themed findings from the identified data in the field. For privacy, we anonymize the usernames used in Subreddit discussions. 

\subsection{Conflicting voices}

\paragraph{Telling stories is less attractive than making money.}
As previously mentioned, in the first set of postings we identified, each discussion thread started with a request from a journalist or researcher for an interview with Turkers. The first reply from Turkers to each request seemed to be not just subtly uncooperative, but also to convey a primary concern with financial compensation when being requested to perform a task for free that they might otherwise earn money from. Some examples of such first replies include:

\begin{quote} 
\hspace{1em}\emph{“Want work from a turker? Post a HIT on the platform.”} [Turker1] 

\hspace{1em}\emph{“If you really want turker's opinions, you could set up a hit for it I'm sure.”} [Turker2]

\hspace{1em}\emph{“How much does this HIT pay?} [Turker3]

\hspace{1em}\emph{“Yeah, a lot of us posted wanting to know what you're paying. We measure our lives by the minute.”} [Turker4]

\hspace{1em}\emph{“What's the pay/time commitment? I used to be a decent guy until I realized that an hour of time can mean the difference between being late on the electric bill.”} [Turker5]
\end{quote}

These initial responses showed no interest in sharing their working stories. Instead, they seemed to view such requests as labor and equivalent to the paid tasks they perform on AMT. Turker4 and Turker5 put their financial consideration in tangible terms: for them, spending time to respond to such a request would take time away from completing more HITs. To some extent, their comments also make a case for the immediate financial gain of microwork -- "an hour of time" can make a difference in their earnings. In contrast, an opportunity to share their stories appears to be uncompelling, at least in the context of these specific requests and their intended usage. On various occasions, we observed some Turkers sharing their prior experience with such interview requests, suggesting that it is not an uncommon event in this community.

\begin{quote} 
\hspace{1em}\emph{“Some reporter shows up in this sub}[Reddit] \emph{about once every 3-4 months soliciting workers for free info. If you want info about our work, make a HIT and pay us for our time. You're literally coming to our forum and asking us to do the type of thing we do on MTurk, but for free.”} [Turker6]

\hspace{1em}\emph{“Bummer, I just did a 70 minute interview with the New York Times for an article that's coming out at the end of the week. I didn't get paid for it though.”} [Turker7]
\end{quote}

In every thread we analyzed, we rarely found that Turkers showed enthusiasm or expressed positive reactions to the interview requests. Instead, we consistently observed \say{negative} experiences of Turkers as to these interview requests in the forum; workers had done one but did not feel they were adequately compensated, or they had seen these requests often and were generally not satisfied with being asked to perform labor without compensation for the time involved.

\paragraph{Distrust of media.}
While compensation could be one cause for the rarity of positive reactions from Turkers to these interview requests, some comments reveal more aspects of their negative experiences with interviews:

\begin{quote}
\hspace{1em}\emph{“Twice I've done interviews with a journalist about MTurk and both times they misrepresented what I said. Never again”} [Turker8]

\hspace{1em}\emph{"...I have yet to see an article that doesn't quote people out of context and/or distort the facts entirely just to suit the title of the article."}[Turker9]

\hspace{1em}\emph{"With all due respect, the last few times there have been magazine articles, it has not gone well. There is always a promise to tell our side, and "give us exposure". The reality ends up being a smear campaign. Here are some examples..."}[Turker10]

\hspace{1em}\emph{"Be careful what you write my friend. With the declining state of print media and paid journalism you'll be one of us before too long."}[Turker11]
\end{quote}

These accounts reflect self-awareness of how Turkers are represented in media narratives. In the elided portion of the quotation above, Turker10 listed three articles from well-known media sources as examples of what they meant by \say{a promise to tell our side} but ending up with \say{being a smear campaign}. In that particular thread, the journalist followed up, referring to existing academic work that demonstrates some visibility is needed for improving Turkers' working condition. However, Turkers in that thread did not seem to appreciate the work. 

One Turker asserted that they would trust their own experience rather than a theory written by others. Another Turker, more pointedly, alleged that both academic work and journalist articles \emph{\say{utilized the lowest tier of workers to create some sort of sob story that all mturk workers are being taken advantage of}}, and these works are \emph{\say{poorly perceived in the turking community}}. Turker11 wrote with a sarcastic tone. We observe distrust toward the journalists and researchers who post, due to Turkers being disappointed by prior interviews and/or news articles and concerned with distortion and misrepresentation. 

In Turkers' reactions to news articles we reviewed, this sentiment seems to be salient: they tend to express a low expectation for any coverage up-front and only react more positively when find the article does not paint a hopeless picture of them. For instance, in reaction to recent coverage from a TV channel, one Turker posted:

\begin{quote}
\hspace{1em}\emph{“At least her team did their study calling us Turkers.”} [Turker12]
\end{quote}

Turker12 here does not comment about the news coverage itself, but notes that it is not that bad because they were at least being addressed in the right way, showing no greater expectation for the reporting. Indeed, in some threads of interview requests, when the journalist or researcher addressed Turkers alternatively such as \say{MTurks} and \say{Amazonians}, Turkers reacted with a stronger tone, trying to establish their identity as \say{workers} and sometimes \say{Turkers}. 

\begin{quote}
\hspace{1em}\emph{“First things first: we're not MTurks. MTurk is the platform. We are workers.”} [Turker13]

\hspace{1em}\emph{"Mturks": the FIRST CLUE you've never even so much as looked at the platform, because if you did, you'd know we're WORKERS (meaning, people).”} [Turker14]

\end{quote}

In another discussion thread, where Turkers commented on a recent media article about them, concerns expressed showed that workers were apprehensive even before reading the article or became worried when they saw the headline or the first few statements of the article solely emphasizing low pay. However, after reading the full article, upon finding that the news story did not just present a one-sided account of crowd work, workers often expressed relief. 

\paragraph{Fallacies of representation.}
Another Turker had more specific comments in the aforementioned discussion about the TV coverage of Turking, pointing to two potential fallacies in news representation of Turkers: first, who from the Turker population was being interviewed; second, whether the news producers had a formulated narrative beforehand that could lead to selective reporting. 

\begin{quote}
\hspace{1em}\emph{“The problem with the articles she cited is that they're either interviewing people who are ridiculously new to the platform and just view it as "\$5/day beer money" or whatever, or they're stacked by people who have an agenda and seek to push a specific narrative...”} [Turker15]
\end{quote}

In the comments on different media articles, we found similar voices alleging that there is a pre-set agenda in news reporting about Turkers, particularly about their status of being \say{underpaid and overworked and exploited}. Some Turkers in the forum expressed fatigue regarding this recurrent narrative.

\begin{quote}
\hspace{1em}\emph{“...the very studies that push the "underpaid and overworked and exploited" were all written with an agenda in mind. Instead of researching what would be a fair pay for turkers, some requesters just up and leave or post their work on other platforms.”} [Turker16]

\hspace{1em}\emph{“...Most MTurk communities have seen this article rehashed time and time again, and we all knew what this was going to say long before it came out.”} [Turker17]
\end{quote}

As we examined the original media coverage being discussed in our identified forum threads, we also noticed common analogies repeatedly used in media coverage to describe crowd work. For example, in one discussion thread, Turkers appear in a media article \cite{lim_why_2018} that refers to gig work as \say{digital slavery} and crowd work as \say{click farms}. Turkers expressed their unease:

\begin{quote}
\hspace{1em}\emph{“I find the "slavery" analogies offensive…”} [Turker18]

\end{quote}

Under another article posted that frames crowd work platforms as \say{virtual sweatshop}, one Turker replied sarcastically:

\begin{quote}
\hspace{1em}\emph{“"digital underclass" is the best description I've heard of turking so far.”} [Turker19]

\end{quote}

\paragraph{Anxiety about publicity.}
When Turkers exhibit resistance to interview requests and passivity to media coverage, they demonstrate unease with the possible or gained publicity. For example, when expressing some dissatisfaction with some media reporting, a Turker voiced clear opposition to such publicity:

\begin{quote}
\hspace{1em}\emph{“Below, you will find three examples of why this will be a hard pass from almost everyone who has turked for a while. We are fine without extra "help" or "publicity”} [Turker20]
\end{quote}

Turker20 then commented on three articles that have given unfavorable publicity to Turkers. The reason that Turker20 held was that to portray Turkers as \say{exploited} by digital platforms could turn task Requesters away because crowd work would be perceived as \say{unethical}. However, Turker20 also acknowledged that even in the case of positive publicity about Turking, it would just lead to more workers coming to the platform, competing with existing Turkers for limited task offerings. Similar voices abound:

\begin{quote}
\hspace{1em}\emph{“Please don't give mturk any more publicity. There is a limited amount of work available, and if people learn about mturk and decide to join, it's less money for us”} [Turker21]

\hspace{1em}\emph{“Publicity really isn't appreciated at this juncture. We're just coming away from a rough NYT article and I think most turkers wish the 4th estate} [press and news media] \emph{would forget about us for a little while.”} [Turker22]

\hspace{1em}\emph{“Talking to journalists is literally putting your income at risk for no reason.”} [Turker23]
\end{quote}

Essentially, Turkers would rather have their Turking opportunities secured, instead of gaining visibility that could jeopardize their opportunities for earning income. Their sentiment expresses risk aversion, since there is also the possibility that negative publicity could drive current Turkers away from the platform (thus fewer competitors for existing Turkers), and positive publicity could bring more Requesters to the platform. Potential downsides occupy their focus.

\subsection{Diverse voices}
\paragraph{The omitted diversity.}
In one case when a journalist tried to recruit workers to interview, the journalist even promised that \say{this isn't going to be another digital sweatshop story}. One worker still responded:

\begin{quote}
\hspace{1em}\emph{“The problem with these journalists is that they just go around and PM random users on this sub[reddit] and just get totally wildly different answers. Mturk is very subjective, some people are terrible at it and/or don[']t use scripts and might only make \$4 a day and meanwhile someone who really gets Mturk and/or uses scripts can make \$40 a day in the same amount of time as someone who doesn[']t use scripts.”} [Turker24]

\end{quote}

This quotation speaks of some reasons why Turkers are not represented properly in media coverage. The experience of Turking varies wildly from person to person. Excess attention given to a universal, powerless image of Turkers in the media narratives hides some Turkers who are \say{motivated self-starters}, as spoken of by a Turker in the forum. This image overshadows the fact that microwork also requires skills, expertise, and perseverance. While existing academic work has highlighted skills and strategies for Turking (e.g., \cite{savage_becoming_2020, kaplan_striving_2018}), we see here, as one Turker put it: 

\begin{quote}
\hspace{1em}\emph{“But I will say this, the majority of high earners usually don't browse this sub or bother with this subreddit because they are too busy working on making \$100+ a day and just see the people here as fools (for the lack of a better word).”} [Turker25]
\end{quote}

It could be possible that high-earning Turkers have less visibility in this Subreddit, from which our analysis is drawn. In the same thread, another Turker noted that success in Turking requires dedication and perseverance. These qualities are for learning the nitty-gritty of the platform, closely observing and acting with the dynamics of the task marketplace, and strategically making use of Turking tools. Prior work \cite{gupta_turk-life_2014,savage_becoming_2020, hanrahan2021expertise} has noted that there is a ladder of experience levels among Turkers, from \say{novice} to \say{expert}. It is likely that those aspects are given far less attention in the popular narratives about Turking \cite{chris-turkerview}.

\paragraph{Diverse motivations for Turking.}
Apart from varying experience levels, motivations for Turking also vary greatly. We observed that some Turkers indeed spoke out about their own conditions and experiences. Consistent with the findings above, these were not early responses to interview requests nor reactions to media articles. We saw that after a post develops into a longer discussion, some Turkers begin to open up more about their individual circumstances. In such accounts, across threads, Turkers shared their diverse motivations for Turking. Some described themselves as socially anxious, autistic, or lacking the competence to thrive in a regular labor market. Turking was characterized as providing them with an alternative mode of work. Some workers expressed appreciation for the flexibility of working outside an office or simply the pleasure of incremental achievements. Some Turkers were explicit about doing \say{side gig} for making use of their idle time to earn \say{guilt-free money} to treat themselves. 

\begin{quote}
\hspace{1em}\emph{“I don't accept HITs that pay that low. I make around \$10/hour and I use that money for my bubble tea habit, Christmas presents for family and generally being able to treat myself occasionally.”} [Turker26]

\hspace{1em}\emph{“Sometimes when I drop my daughter off for her 3-4 hour volleyball practices, rather than go home and sit around I go to Starbucks and turk over a cup of coffee or two.”} [Turker27]

\hspace{1em}\emph{“I’ve paid entirely for a high end camera with Turk money, and this has become my favorite hobby and returned priceless memories for my family.} [Turker28]

\hspace{1em}\emph{“My ability to do Turking adds \$400-\$600 a month without taking me out of my primary role, which means nearly the same amount of income as her previous position with much more family time.”} [Turker29]
\end{quote}

In these posts, Turking sounds rewarding and enjoyable to Turkers. In contrast, some workers report reliance on Turking for their primary income:

\begin{quote}
\hspace{1em}\emph{“I do it because I'm not healthy enough to handle a "real" job right now. Turking lets me work from home whenever I feel able and take as many breaks as I need. ”} [Turker30]

\hspace{1em}\emph{“I am a mom of three children and am trying to bring in extra money with MTurk. I have a degree but am unable to work outside of the home because two of my children have significant special needs.”} [Turker31]

\hspace{1em}\emph{“I'm physically not able to go out to work right now, but I still need food and shelter. I'm not officially disabled (no health insurance = no doctor) so I'm not getting any money from the government. I'm doing as many online jobs as I can right now, but those aren't so plentiful. I turk to survive.”} [Turker32]

\end{quote}

Prior work \cite{antin_social_2012, ipeirotis2010demographics, kaufmann_more_2011, kaplan_striving_2018} has documented a wide range of motivations for Turking across nationality and geography.
From these accounts we collected in the forum, Turkers' motivations and perceptions of Turking vary according to their life conditions. One is naturally less likely to view Turking as entertainment when it serves as one's primary source of income.

\paragraph{Diverse career perspectives.}
Turkers come to MTurk with different life conditions and attitudes. As we saw in the Subreddit, some workers do not think Turking to be a \say{job} or a \say{career}, bringing correspondingly lower expectations for the work. In contrast, other dedicated Turkers in the forum contend that:

\begin{quote}
\hspace{1em}\emph{“I treat my work on here like a real job and I guess many other}[s] \emph{do not.”} [Turker33]

\hspace{1em}\emph{“If you look at it like a job...then you are getting "training" or learning what that "boss" wants. It isn't worth it sometimes if there is only one hit...but if it's something you can bank like you just mentioned, it's worth it.”} [Turker34]

\hspace{1em}\emph{“I don[']t even try and argue with them anymore. More work for me and others who diligently read when they cry and complain.”} [Turker35]

\end{quote}

Turker33 thinks the view of Turking as \say{a real job} is not common among Turkers. But Turker34 adopts a serious attitude toward Turking as a job and recognizes the personal growth gained through Turking. One Turker from the same thread comments that \say{the learning curve is steep} for crowd work while another affirms that to make the platform work for oneself \say{time, experience, and a certain mindset} are required. Turker35 expresses a very practical attitude that it is not worthwhile to get upset or argue with task Requesters who reject their work, while outrage regarding unfair rejection often features prominently in popular press articles about Turking.

As we understand from these online discussions, recognizing Turking as formal career experience seems helpful to breed resilience. These accounts suggest that Turking is seen as a formal career option for some workers, in contrast with the popular media portrayal of an exploited workforce.

\section{Discussion}
Based on our analysis of online discussion data, we have seen that in the online Turker discourse community, Turkers tend to be disinclined to accept interview requests. Specifically, Turkers are resistant to the idea of being interviewed and sharing their stories, as they fear that publicity will turn their work environment more competitive. They do not applaud media narratives that portray them as a powerless and exploited underclass, even if their working condition is not ideal. 

In general, Turkers seem averse to media coverage adopting a more \say{activist} tone as a preferable pathway to improve their working status. They express a stronger attitude of self-determination and self-reliance regarding their working decisions and whether or not anything needs to be rectified, and how to go about it if so. They emphasize a simple desire for secure income available from Turking. 

Many researchers have discussed the invisibility of crowd workers (e.g., \cite{irani_turkopticon_2013, gray_ghost_2019}), and how this can lead to  misrepresentation of the work involved and increase precarity of work \cite{suchman_making_1995, star_layers_1999}. Many academic efforts seek to probe into the invisible crowd work and gain a deeper understanding. While online Turker forums could serve as a channel for understanding these communities from outside an opaque platform, there exist additional layers of friction to the elicitation of Turkers' voices and accounts. Turkers do not just easily tell their stories. 

In order to better understand Turkers' experiences so that we might devise methods to better support them and their work, we need to better understand and respect their experiences and wishes. Well-meaning but naive or poorly-executed interventions can cause harm and provoke anger from the very people whom such interventions seek to help. While various studies have identifies problems with worker invisibility, simply giving Turkers more exposure can induce stress and resistance. Turkers largely do not seem to want the attention that may disrupt their work and work environment, though they also express a desire that the platform and income opportunities can be further secured or bolstered. 

One-sided media (or research) accounts of Turkers' stories seem particularly problematic, including distorting the public understanding and narrative around crowd work. Other researchers have also noted this \cite{moss2020ethical}. Many Turkers in the online forum are aware of the simplistic caricatures of crowd work illustrated in popular narratives. In contrast, their own accounts of Turking are often a diverse and mixed bag of upsides and downsides, not necessarily converging into one universal image. There is a clear gap between typical media portrayals of crowd work and workers' own perception of Turking. Use of terminology such as \say{digital slavery} is particularly problematic as misleading, demeaning, and racially charged. Turking itself requires special skills and many HITs are also knowledge-intensive. Turkers' work and professionalism merit our respect. 

When it is difficult to access an opaque platform and elicit Turkers' voices, journalist practices are particularly susceptible to the trap of \say{parachute journalism} \cite{wizda_parachute_1997}. This refers to the production of quick, one-dimensional news pieces that merely capture the clich\'es and stereotypes, often with stories pre-defined before the reporter hits the ground in the locale to be reported on. Some conditions, such as insufficiency of local knowledge and tight deadlines, provide fertile ground for \say{parachute journalism}. Rather than questioning professional ethics, we suggest more attention be given to the nature of crowd work as a complex sociotechnical issue. Less visible online work requires updated practices to learn about its operation in depth. Merely interviewing a few crowd workers is unlikely to capture the rich diversity of lives and lived experiences surrounding crowd work. Several times in worker discussions we saw a critique of whether a journalist's experience on AMT for a couple of weeks is sufficient to understand Turking themselves, let alone depict it accurately to their readers. 

When we write \say{Turkers}, it is tempting to assume this name envelops a homogeneous group, rather than recognizing it as an umbrella term encompassing a rich diversity of underlying humanity and experiences. Beyond well-known geographic diversity, there is diversity in a multitude of other respects, including living conditions, motivations, valuation of Turking, skill and experience levels, and career visions \cite{andy_baio_faces_2008, oppenlaender2020crowd}. Consequently, it is  challenging to effectively support this diverse workforce via any one-size-fits-all intervention, policy, or regulation. 

Because our findings are drawn from online observations in Turker forums and their discussion data, our analysis is limited by that observed data.
We miss the voices of Turkers who do not post in online forums, the forum we selected, or in other threads excluded from our sample. That said, our findings complement the prior research work by shedding further light upon \emph{why} understanding and representing crowd workers is so difficult. Their size, scale, diversity, invisibility, and resistance to publicity make it particularly difficult to access them and risk bias-prone representations. These constitute clear challenges to recognize in pursuing a worker-centered agenda to support crowd work. It is also critical to strive to understand the constituency of Turkers we wish to aid in such work.

For a worker-centered research agenda, what approaches would best enable us to understand the shifting and diverse population of Turkers? How should we respond to the invisibility of the workforce when they are resistant to publicity? How could we achieve a worker-centered approach to crowd work when it is inherently hard to realize so with traditional approaches? The voices of Turkers we have heard lead us to these questions to help guide future research.

\section*{Acknowledgments}
We thank all those who provided feedback to us in preparing this manuscript. This work was supported in part by Good Systems\footnote{\url{https://goodsystems.utexas.edu}}, a UT Austin Grand Challenge to develop responsible AI technologies. Matthew Lease is also supported as an Amazon Scholar. Our opinions are entirely our own and do not speak to those of the sponsoring agencies. 

\bibliographystyle{unsrt}  
\bibliography{ms}

\end{document}